\documentclass[12pt]{iopart}
\usepackage{graphicx}
\usepackage{amssymb}
\usepackage{dcolumn}
\usepackage{bm}

\begin{document}

\title{Exact solvability, non-integrability,  and genuine multipartite entanglement dynamics of the
Dicke model}

\author{Shu He$^{1}$, Liwei Duan$^{1}$, and Qing-Hu Chen$^{1,2,3}$}

\address{$^1$ Department of Physics, Zhejiang University, Hangzhou 310027, P. R. China}
\address{$^2$ Collaborative Innovation Center of Advanced Microstructures, Nanjing University, Nanjing 210093, China}
\ead{ qhchen@zju.edu.cn}

\begin{abstract}
In this paper, the finite size Dicke model of arbitrary number of qubits is solved analytically in an unified way within extended coherent states.
For the $N=2k$ or $2k-1$ Dicke models ($k$ is an integer),  the $G$-function, which is only an energy dependent $k \times k$
determinant, is  derived in a transparent manner. The regular spectrum is completely and uniquely given by stable zeros of
the $G$-function. The closed-form exceptional  eigenvalues  are also derived. The level distribution controlled by the
pole structure of the $G$-functions suggests non-integrability for $N>1$ model at any finite coupling  in the sense of recent criterion in literature.
A preliminary application to  the  exact dynamics of
genuine multipartite entanglement  in the
finite $N$ Dicke model is  presented using the obtained exact solutions.
\end{abstract}

\date{\today}

\pacs{03.65.Fd, 42.50.Ct, 64.70.Tg}

\maketitle

\section{Introduction}

The Dicke model~\cite{dicke} describes the interaction of a number ($N>1$)
of two-level atoms (qubits) with a single bosonic mode and has been a
paradigmatic example of collective quantum behavior. It is closely related
to many fields in quantum optics, quantum information science, and condensed
matter physics. The superradiant behavior by an ensemble of quantum dots~%
\cite{Scheibner}, Bose-Einstein condensates~\cite{Schneble,newexp}, and
coupled arrays of optical cavities has been used to simulate  the
strongly correlated systems~\cite{Hartmann}.

Quantum integrability is an often-mentioned concept in the
description of a quantum system, however it is not well defined to
date~\cite{quntumint}, in sharp contrast with the classical
integrability, mostly due to the markedly different ways to count
the degrees of freedom (d.o.f) in the quantum and classical
mechanics. This situation becomes more serious for the Dicke model,
where the classical limit is lacking because of  the quantum nature
of the finite number of levels, one cannot define integrability from
any classical limit. In the semi-classical limit, where the boson
mode is treated as a classical field while the discrete level is
kept as a quantum entity, the signatures of chaos emerge in the
Dicke model~\cite{nonint}, even in the Rabi model (single
qubit)~\cite{muller}. One appealing way to address the quantum
integrability is looking at the energy level statistics. The
corresponding Berry-Tabor criterion~\cite{BT} states that a quantum
system is integrable if the level statistics is Possionian. However
it relies on the semiclassical arguments and only concerns quantum
systems with continuous d.o.f. The level statistics of the finite
but large $N$ Dicke model is shown to be Possionian below and
Wigner-Dyson above the critical coupling~\cite{quntnoint,Emary}.
While the level statistics of Rabi model is neither Poissonian nor Wigner~%
\cite{kus}, which is characteristics of genuinely integrable models.
Recently, Braak proposed that if the eigenstates of a quantum system
can be uniquely labeled by $f=f_1+f_2$ quantum numbers, where $f_1$
and $f_2$ are the numbers of the discrete and continuous d.o.f, then
it is integrable~\cite {Braak}. By this novel criterion,  the $N=3$
Dicke model is non-integrable~\cite {Braak2013}, while the Rabi
model is integrable\cite {Braak}. Implications of the
(non)-integrability in the Dicke model with arbitrary number of
qubits should be generally interesting.

Despite the simplicity of the full Dicke model, its solution  is however
highly non-trivial. Many approaches have been developed and extended to the
this model~\cite
{Emary,oster,Carollo,Lambert1,Buzek,Lambert2,liberti,vidal,reslen,plastina,
chen2008}. Among them, the exact eigensolutions in the finite $N$ Dicke model are
very crucial to get some deep insights, but they are mainly limited to
numerical diagonalization in truncated Bosonic Fock space~\cite
{Emary,Lambert1,Lambert2}. Recently, Chen \textit{et al.}~\cite{chen2008}
have presented numerical exact solutions to this model using extended
coherent states (ECS)~\cite{chen946}, where the truncation of the Hilbert
space can be alleviated systematically. It has been extensively shown in
Refs.~\cite{Bastarrachea,Hirsch1,Hirsch2} that as compared with the photonic
number (Fock) basis, ECS is exhibited to be valid for a large region of the
Hamiltonian parameter space by analyzing the converged energy eigenvalues
and eigenfunctions.

More recently, an analytical exact solution, which is mathematically well
defined, to the Rabi model~\cite{Rabi} has been discovered by Braak~\cite
{Braak} using the representation of bosonic creation and annihilation
operators in the Bargmann space of analytical functions~\cite{Bargmann}. A
so-called $G$-function with a single energy variable was derived yielding
exact eigensolutions. Alternatively, using ECS, this $G$-function was
recovered in a simpler, yet physically more transparent manner by Chen
\textit{et al.}~\cite{Chen2012}. The continuous extensions to the analytical
treatments for the Dick-type model with less than $3$ qubits have also
achieved progress~\cite
{Braak2013,Zhong,Chilingaryan,zhangyz,Moroz,Rodriguez,Peng,Shen,wang2014,Zhang,Chennext}%
. Among them, complicated $6\times 6$ determinant $G$-functions were derived
in the Bargmann representation for the $N=3$ Dicke model~\cite{Braak2013}. By using the ECS,
$G$-function resembling the most simple one without a determinant in
the  Rabi model was given for the $N=2$ Dicke model~\cite{wang2014}  and $2\times 2$ determinant $G$-function
was derived in the  Rabi model with two arbitrary qubits~\cite{Chennext}. A generally concise analytic solutions to the
Dicke model of arbitrary number of qubits, which is a counterpart to the
previous numerical exact solution~\cite{chen2008}, should be highly called
for.

On the other hand, quantum correlations among qubits, viewed as quantum
information resources, have also attracted extensive interest. A dozen
qubits are sufficient in the applications to quantum information technology.
However, only bipartite quantum correlations have been studied in the Dicke
model~\cite{Lambert1,Buzek,Lambert2,liberti,vidal,reslen,Wang2012}, and
multi-partite quantum correlations have not been explored, to the best of
our knowledge. The multi-partite quantum correlations can be used to
implement operations in the measurement-based quantum computation~\cite
{Briegel} and high-precision metrology~\cite{Giovannetti}, thus is most
important from the experimental viewpoint. However, it was difficult to
characterize multi-partite quantum correlations in many previous efforts~
\cite{multi}. Recently, the theory of multipartite entanglement has been
achieved much progress, especially the computable entanglement monotone~\cite
{Jungnitsch}. On the experimental side, multi-qubits have been constructed
in several solid devices recently, such as the well-known
Greenberger-Horne-Zeilinger (GHZ) states~\cite{Greenberger} of more than two
qubits using the complete circuit and full set of gates~\cite
{DiCarlo,Barends,Wang}. The path to scalability is prerequisite to realize
the quantum computation and quantum information processes.

The goal of this paper is twofold. First, we demonstrate a
successful derivation of a concise $G$-function for the arbitrary
$N$  Dicke model for by means of
ECS~\cite{Chen2012,wang2014,Chennext}. For $N=2k-1$ or $2k$, the
$G$-function is a $k\times k$ determinant, which are very feasible
to calculate all eigensolutions for the Dicke model with a dozen
qubits. The non-integrability of the finite $N$ Dicke model is then
discussed in terms of the level distribution dictated by the
$G$-functions. Second, using the exact eigensolutions, we study the
dynamics of the genuine multipartite entanglement (GME) from the
maximum entangled states, such as Bell states for $N=2$ and GHZ
states for $N=3$ and $4$.

The paper is organized as follows. In Sec. 2, we introduce the model
Hamiltonian and briefly review the previous numerical exact
techniques based on the ECS. Section 3 describes the detailed
analytic scheme for the solutions to the finite $N$ Dicke model. The
$G$-functions are derived explicitly. In Sec. 4, we illustrate that
the exact spectra can be easily obtained by the stable zeros of the
$G$-function and discuss the implications of the present derivation
with the non-integrability of the finite $N$ Dick model. As an
example of the applications of the present method, the exact
dynamics of the GME of the Dicke model is studied using the obtained
exact eigensolutions in Sec. 5. A summary is given in the final
section.

\section{Model}

The Hamiltonian of the finite $N$ Dicke model can be written as ($\hbar =1
$)
\begin{equation}
H=-\Delta J_x+{\omega }d^{\dag }d+2\lambda /\sqrt{N}\;(d^{\dag }+d)J_z,
\label{Hamiltonian}
\end{equation}
where $\Delta $ is the qubit splitting, $d^{\dag }$ creates one photon in
the common single-mode cavity with frequency $\omega $, $\lambda $ describes
the atom-cavity coupling strength, (we usually denote$\;g=\lambda /\sqrt{N}$%
), and $J_z=\frac 12\sum_i\sigma _z^{(i)},J_x=\frac 12\sum_i\sigma
_x^{(i)}=\frac 12\left( J_{+}+J_{-}\right) \;$where$\ $ $\sigma _{x,z}^{(i)}~
$ is the Pauli operator of the $i$-th qubit, $J$ is the usual angular
momentum operator, and $J_{+}$ and $J_{-}$ are the angular raising and
lowing operators and obey the $SU(2)$ Lie algebra $%
[J_{+},J_{-}]=2J_z,[J_z,J_{\pm }]=\pm J_{\pm }$. This Hamiltonian posses a $%
Z_2$-symmetry like the Rabi model and exhibits a second-order quantum phase
transition at $\lambda _c=1/2\;$in the thermodynamic limit \cite{Emary}. In
this work, the Hilbert space is spanned by the Dicke state $\left| \frac
N2,m\right\rangle $, which is the eigenstate of $J^2$ and $J_z$ with the
eigenvalues $\frac N2(\frac N2+1)$ and $m$. Throughout this paper, the unit
is taken of $\ \omega =1$.

Previously, we have proposed a numerical exact approach for Dicke model of
arbitrary $N$ by ECS~\cite{chen2008}. The wave function is described in the
basis $\left| j=\frac N2,m\right\rangle \bigotimes \left| \varphi
_m\right\rangle _d$ where the photonic state is given by
\begin{equation}
\left| \varphi _m\right\rangle =\sum_{k=0}^{N_c}u_{m,n}\frac 1{\sqrt{n!}%
}(d^{\dag}+g_m)^ne^{-g_md^{\dag}-\left( g_m\right) ^2/2}\left|
0\right\rangle , \label{wavefunction}
\end{equation}
where $\left| 0\right\rangle $ is the photonic vacuum state, $%
g_m=2gm\;\left( m=j,...,-j\right) ,$and $N_c$ is the truncated bosonic
number in the Fock space of the new operator $A_m^{\dag }=d^{\dag }+g_m$.
The coefficient $u_{m,n}$ can be determined through the exact Lanczos
diagonalization. In that work, we have solved the Dicke model for more than
one thousand qubits numerically, and  obtained some critical exponents
with high accuracy. It was numerically shown later that increasing the
number of atoms imposes a strong limit to the states in the Fock basis, but
not in the ECS scheme~\cite{Bastarrachea,Hirsch1}. Interestingly, the
similar or same ansatz for the wavefunction like (\ref{wavefunction}) have
been used in many recent works for the Dick-type models.

In the next section, we will propose an analytic scheme to the exact
solution for the finite  $N$ Dicke model.

\section{Analytical scheme to exact solutions}

We will demonstrate that the $N=2k-1$ and $N=2k$ Dicke models can be treated
in an unified way. In the matrix form of the Dicke state basis, the
Hamiltonian takes a tridiagonal type. The diagonal element is $%
H_{m,m}=d^{\dag }d+2mg(d^{\dag }+d)$. While  two typical
off-diagonal elements are $H_{m,m+1}=-\frac \Delta 2j_m^{+}\left(
1-\delta _{2m,N}\right)$ and $H_{m,m-1}=-\frac \Delta 2j_m^{-}\left(
1-\delta _{2m,-N}\right) $, where $j_m^{\pm }=\sqrt{\frac N2(\frac
N2+1)-m(m\pm 1)}$ and $\delta $ is the Kronecker delta.

We introduce $k$ pairs of displaced operators as
\begin{equation}
A_{\pm m}^{\dag }=d^{\dag }\pm 2\left| m\right| g, \label{displaced}
\end{equation}
where $m=0$ for $N$ even case is excluded here, because it is just the original photonic
operators without displacement and will be independently treated later.
By $A_{\pm m}^{\dag }$,
the diagonal element $H_{m^{\prime },m^{\prime }}$ becomes
\begin{equation}
H_{m^{\prime },m^{\prime }}=A_{+m}^{\dag }A_{+m}+2\left( m^{\prime
}-m\right) g\left( A_{+m}^{\dag }+A_{+m}\right) -4m\left( 2m^{\prime
}-m\right) g^2.  \label{H_mm_n}
\end{equation}
Specially, the diagonal element $H_{m,m}$ for $m^{\prime} =m$ can be
transformed into the form with free particle number operators.

Then we propose the wavefunction in Fock space of $A_{+m}^{\dagger }\;$
\begin{equation}
\left| A_{+m}\right\rangle \varpropto \sum_{m^{\prime
}=-j}^j\sum_{n=0}^\infty \sqrt{n!}c_{m^{\prime },n}^{(m)}\left| \frac
N2,m^{\prime }\right\rangle |n\rangle _{A_{+m}},  \label{wave_A}
\end{equation}
where $c_{m^{\prime },n}$ are coefficients and $|n\rangle _{A_{+m}}$ are the
number states in $A_{+m}^{\dagger }$. The latter is just called as ECS ~\cite{Chen2012} with the
following property
\begin{equation}
|n\rangle _{A_{+m}}=\frac{\left( d^{\dagger }+2mg\right) ^n}{\sqrt{n!}}%
|0\rangle _{A_{+m}};\;|0\rangle _{A_{+m}}=e^{-2m^2g^2-2mgd^{\dagger }}\left|
0\right\rangle _d,
\end{equation}
where the vacuum state $|0\rangle _{A_{+m}}$ is the eigenstate of
the one-photon annihilation operator $d$ with eigenvalue $-2mg$.

The Schr\"{o}dinger equation leads to the recurrence relation for the
expanded coefficients as
\begin{eqnarray}
c_{m^{\prime }\neq m,n+1}^{(m)} &=&\frac{E-\left[ n-4m\left( 2m^{\prime
}-m\right) g^2\right] }{2\left( m^{\prime }-m\right) \left( n+1\right) g}%
c_{m^{\prime },n}^{(m)}-\frac{c_{m^{\prime },n-1}^{(m)}}{n+1}  \nonumber \\
&&+\frac{H_{m^{\prime },m^{\prime }+1}c_{m^{\prime }+1,n}^{(m)}+H_{m^{\prime
},m^{\prime }-1}c_{m^{\prime }-1,n}^{(m)}}{2\left( m^{\prime }-m\right)
\left( n+1\right) g},  \label{recur_m}
\end{eqnarray}
and
\begin{equation}
c_{m,n}^{(m)}=\frac{H_{m,m+1}c_{m+1,n}^{(m)}+H_{m,m-1}c_{m-1,n}^{(m)}}{%
E-\left( n-4m^2g^2\right) }.  \label{recur_n}
\end{equation}
We can see that $c_{m,0}^{(m)}$ can be determined by $c_{m+1,0}^{(m)}$ and $%
c_{m-1,0}^{(m)}$, so all energy $E$ dependent coefficient $c_{m^{\prime
},n}^{(m)}$ can be determined by $N$ initial parameters $c_{m^{\prime }\neq
m,0}^{(m)}$ recursively.

The Hamiltonian (\ref{Hamiltonian}) is invariant under the transformations $%
m\leftrightarrow -m\;$and $d^{\dag }\left( d\right) \leftrightarrow -d^{\dag
}\left( -d\right) $, due to the associated conserved parity. So the
wavefunction in the series expansion in $A_{-m}^{\dagger }$ should take
\begin{equation}
\left| A_{-m}\right\rangle =\sum_{m^{\prime }=-j}^j\sum_{n=0}^\infty (-1)^n%
\sqrt{n!}c_{-m^{\prime },n}^{(m)}\left| \frac N2,m^{\prime }\right\rangle
|n\rangle _{A_{-m}}.  \label{wave_A-}
\end{equation}
For the same non-degenerated states, $\left| A_{+m}\right\rangle $ should be
proportional to $\left| A_{-m}\right\rangle $. Comparing the Dicke states $%
\left| j,m\right\rangle $ and $\left| j,j-m\right\rangle $ gives
\[
\sum_{n=0}^\infty \sqrt{n!}c_{\pm m,n}^{(m)}|n\rangle
_{A_{+m}}=r\sum_{n=0}^\infty (-1)^n\sqrt{n!}c_{\mp m,n}^{(m)}|n\rangle
_{A_{-m}}.
\]
Left multiplying $\langle 0|$ and with the use of
$\sqrt{n!}\left\langle 0\right| |n\rangle
_{A_{+m}}=(-1)^n\sqrt{n!}\left\langle 0\right| |n\rangle
_{A_{-m}}=e^{-2m^2g^2}(2mg)^n$, we have one linear equation for one
pair of the displaced operator $A_{\pm m}^{\dag }$,
\begin{equation}
G_{\pm }^{\left( m\right) }=\sum_{n=0}^\infty \left[ c_{m,n}^{(m)}\mp
c_{-m,n}^{(m)}\right] (2mg)^n=0,  \label{G_function_n}
\end{equation}
where $+(-)$ in the l.h.s denotes the corresponding states of positive
(negative) parity.

Therefore we have $k$ sets of linear equations for $k$ pairs of
displaced operators. For the most simple  Rabi model, we only have
one linear equation with either parity, and the relevant
coefficients $c_{\frac 12,n}^{(\frac 12)}$ and $c_{-\frac
12,n}^{(\frac 12)}$ can be determined recursively from the initial
value $c_{-\frac 12,0}^{(\frac 12)}=1$~\cite{Chen2012}, because Eqs.
(\ref{recur_m}) and (\ref {recur_n}) in this case can be reduced to
a linear three-term recurrence relation. For $N > 1$, such linear
three-term recurrence relations are not available, and the
coefficients cannot be obtained from a single initial one, but
instead, from  $N$ initial parameters for each $k$. Then one may
naively think that all coefficients in the $k$ sets of linear
equations (\ref{G_function_n}) should be determined by $k\times N$
initial parameters. Fortunately, we will show below that  $k$
independent initial parameters  would determine all coefficients
 recursively.

The wavefunction can be also expressed in original Fock space as
\begin{equation}
\left| d\right\rangle =\sum_{m^{\prime }=-j}^j\sum_{n=0}^\infty \sqrt{n!}%
a_{m^{\prime },n}\left| j,m^{\prime }\right\rangle |n\rangle _d,
\label{wave_d}
\end{equation}
where $a_{-m^{\prime },n}=\pm \left( -1\right) ^na_{m^{\prime },n}$
with  $+(-)$ positive (negative) parity. Note that this auxiliary
expansion itself exists intrinsically  for $m=0$ in the $N$ even
case, c.f. Eq. (\ref{displaced}).  By the Schr\"{o}dinger equation,
all coefficients $a_{m^{\prime },n}$
can be determined by $k$ initial ones $%
a_{m^{\prime }\neq 0,n=0}$ recursively
\begin{eqnarray}
a_{m^{\prime }\neq 0,,n+1} &=&\frac{E-n}{2m^{\prime }\left( n+1\right) g}%
a_{m^{\prime },n}-\frac 1{n+1}a_{m^{\prime },n-1}  \nonumber \\
&&+\frac{H_{m^{\prime },m^{\prime }+1}a_{m^{\prime }+1,n}+H_{m^{\prime
},m^{\prime }-1}a_{m^{\prime }-1,n}}{2m^{\prime }\left( n+1\right) g},
\label{recur_dm}
\end{eqnarray}
and
\begin{equation}
a_{0,n}=-a_{1,n}\frac{\Delta \left[ 1\pm \left( -1\right) ^n\right] }{%
4\left( E-n\right) }\sqrt{\frac N2(\frac N2+1)}.  \label{recur_d0}
\end{equation}
Note that $a_{0,n}$ is only present for the $N$ even Dicke model.

Similarly, the same non-degenerated states in Eqs. (\ref{wave_A}) and (\ref
{wave_d}) yields
\begin{equation}
\sum_{n=0}^\infty \sqrt{n!}c_{m^{\prime },n}^{(m)}|n\rangle
_{A_{+m}}\varpropto \sum_{n=0}^\infty \sqrt{n!}a_{m^{\prime },n}|n\rangle _d.
\label{relation}
\end{equation}
Projecting onto $_{A_{+m}}\langle 0|$ gives the initial coefficient $%
c_{m^{\prime }\neq m,0}^{(m)}$ in Eq. (\ref{recur_m}) in terms of $%
a_{m^{\prime },n}$ as
\begin{equation}
c_{m^{\prime }\neq m,0}^{(m)}\propto \sum_{n=0}^\infty a_{m^{\prime },n\
}\left( -2mg\right) ^n,  \label{initial}
\end{equation}
where the use has been made of
\[
_{\ A_{+m}}\langle 0|n\rangle _d=\sqrt{\frac 1{n!}}e^{-2m^2g^2}\left(
-2mg\right) ^n.
\]
To this end, through Eqs. (\ref{recur_m}), (\ref{recur_dm}), and (\ref
{initial}) , all coefficients $c_{m,n}^{(m)}$ in Eq. (\ref{G_function_n})
can be determined from $k$ initial coefficients $a_{m^{\prime }\neq 0,n=0}$.
Then we have $k$ sets of linear equation for $k$ unknown variables $a_{m\neq
0,n=0}$.

For the $m$th linear equation, set $a_{m^{\prime },n=0}=1$, and all other
initial variables to be zero, the matrix element $G_{\pm }\left( m^{\prime
},m\right) $ is given by the summation in Eq. (\ref{G_function_n}). The $G$%
-function for the Dicke model is thus defined as the following $k\times k$
determinant
\begin{equation}
G_{\pm }(E)=\det G_{\pm }(m^{\prime },m).  \label{G-function}
\end{equation}
For nonzero $k$ unknown variables $a_{m\neq 0,n=0},$ $G_{\pm }(E)=0$ is
required. The zeros thus give all eigenvalues of the Dicke model, which in
turn give the eigenstates using Eq. (\ref{wave_A}) or Eq. (\ref{wave_d}). It
is interesting to note from the coefficients in Eqs. ( \ref{recur_m}) and (%
\ref{initial}) that the present $G$-function is a well defined
transcendental function. Thus analytical exact solutions have been formally
found for the Dicke model.

\textsl{Exceptional solutions:} Similar to the Rabi model~\cite
{Braak,Chen2012}, for some special model parameters, there are exceptional
solutions that do not correspond to the zeros of the $G$-functions. They can
be also obtained from the pole singularities in Eq.~(\ref{recur_n})
\begin{equation}
E_{ex}^{(m)}=n-4m^2g^2,  \label{exceptional}
\end{equation}
for $m>0$. It is just the $m$th kind exceptional solution. So we totally
have $k$ kinds of exceptional solutions, which are just the eigenvalues in
the atomic degenerate limit ($\Delta =0$). Inserting $E_{ex}^{(m)}$ into the
$k$ sets of linear equations (\ref{G_function_n}), we can note that the
equation corresponding to $A_{+m}^{\dag }$ is not available due to the
singularity, which can be replaced by
\begin{equation}
H_{m,m+1}c_{m+1,n}^{(m)}+H_{m,m-1}c_{m-1,n}^{(m)}=0,
\end{equation}
i.e. the numerator in Eq.~(\ref{recur_n}) vanishes so that the pole is
lifted. By Eq. (\ref{relation}), we have
\[
\sum_{n=0}^\infty \sqrt{n!}\left( H_{m,m+1}a_{m+1,n}+H_{m,m-1\
}a_{m-1,n}\right) \left( -2mg\right) ^n=0.
\]
So there are still $k$ sets of linear equations available for the
$m$th kind exceptional eigenvalue. The zeros of the corresponding
$k\times k$ determinant will yield the condition for the occurrence
of the exceptional eigenvalues. For a given $\Delta $, we can obtain
the coupling constants $g_{\pm }^{n,m}$, which can also be located
by the crossing points of curves for Eq. ( \ref{exceptional}) and
the corresponding energy levels in the spectral graph.  Generally,
$g_{+}^{n,m}\neq g_{-}^{n,m}$, owing to the different parity
dependent conditions.  It follows that exceptional eigensolutions
are non-degenerate, as noted in the $N=3$ Dicke
model~\cite{Braak2013}. It should be stressed here that the level
crossing in two parity subspace does not occur at an exceptional
eigenvalue, and therefore the doubly degenerate state at the level
crossing has the regular spectrum with fixed parity for the $N>1$
Dicke model, in sharp contrast with the Rabi model.

These exceptional solutions result in $k$ kinds of pole structure in
the $G$-function for both the $N=2k-1$ and the $2k$ Dicke models.
For $N=2k$, there exists additional pole at $E=n$ even (odd) for the
positive (negative) parity, as can be seen from Eq. (
\ref{recur_d0}). It is just the eigenvalue for $j=0$ subspace, and
isolated from other Dicke states with non-zero $j$. It is not an
exceptional solution but yields divergence as well, as demonstrated
in the $N=2$ Dicke model \cite{wang2014}. It will be shown later
that all these poles are very helpful to analyze the distribution of
the zeros, namely eigenvalues, in the $G$-function.

\section{Numerical zeros of $G$-functions and non-integrability}

\begin{figure}[tbp]
\center
\includegraphics[width=12cm]{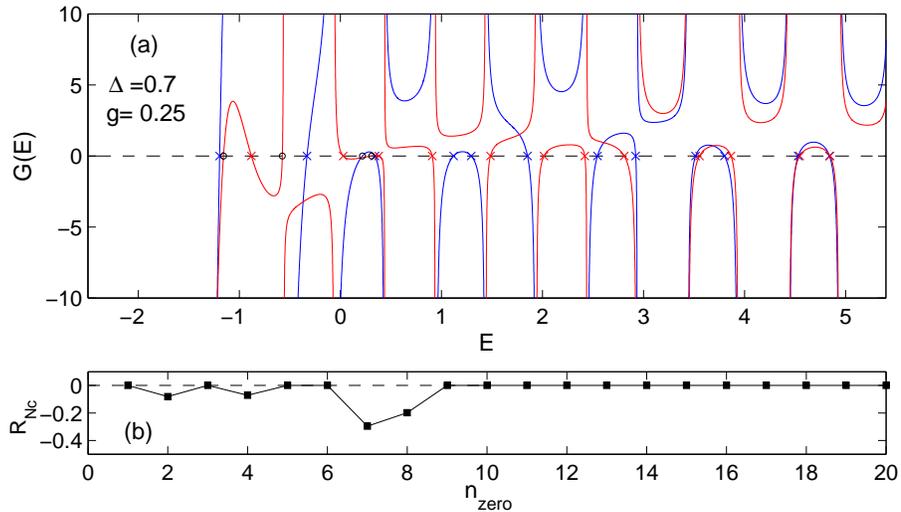}
\caption{ (Color online) (a) G function for the $N=3$ Dicke model with positive
(blue) and negative (red) parity. Crosses denote the zeros. (b) $R_{N_c}=\ln
\left( E_{N_c}/E_{N_c+1}\right) $ for different zeros in (a) with the serial
number $n_{zero}$. $\Delta =0.7$ and $g=0.25$. }
\label{G_function_N3}
\end{figure}

Analogous to the Rabi model~\cite{Braak,Chen2012}, zeros of the $G$
-functions in the Dicke model cannot be obtained analytically in the
closed form either. Searching for the zeros numerically should be
required practically. The curves of $G_{\pm }\left( E\right) $
defined in Eq. (\ref{G-function}) is plotted in Fig. \ref
{G_function_N3}(a) for $\Delta =0.7\;$and$\;g=0.25$ for the $N=3$
Dicke model. The stable zeros give all eigenvalues in the plotted
energy regime, which has been confirmed with the numerical exact
solutions. Typically, the convergence is assumed to be achieved if
zeros (i.e. $E$) are determined within relative errors $\left|
\left( E_{N_c}-E_{N_c+1}\right) /E_{N_c+1}\right| <10^{-8}$, where
$N_c$ is the truncated number $\ $in the
series expansions for the displaced operators. We also calculate $%
R_{N_c}=\ln \left( E_{N_c}/E_{N_c+1}\right) $ for all zeros in Fig. \ref
{G_function_N3}(b). The stable zeros, where $R_{N_c}$ is almost the same as
the relative errors, should be on the $R_{N_c}=0$ line within error $10^{-8}$%
, much smaller than the symbol size.

One can also notice that very few unstable zeros emerge in practical
calculations because of inevitable finite truncations. They do not
belong to the true eigenvalues and can be excluded very easily. They
are very sensitive to the truncated number $N_c$ and cannot converge
with  $N_c$, because the corresponding coefficients $c_{m,n}^{(m)}$
oscillate with increasing magnitudes as $n$ increases. In sharp
contrast, for the stable zeros, the coefficients $c_{m,n}^{(m)}$
converge to zero rapidly with increasing $n$. The positions of
unstable zeros must change if increasing $N_c$ by $1$. So they are
easily figured out in Fig. \ref {G_function_N3} (b) for apparent
deviation from $R_{N_c}=0$ line. It should be pointed out here that
unstable zeros are absolutely not the true zeros of $G$-function if
the summations are really performed infinitely. As demonstrated in
Fig. \ref{G_function_N3} (b) that  their positions shift to higher
energy regime ($E_{N_c+1}>E_{N_c}$) with increasing $N_c$.
Theoretically, the unstable zeros disappear in the finite energy
spectra if $N_c\rightarrow \infty $.

The baselines shown in Fig. \ref{G_function_N3} (a) are close to $m=3/2$ and
$m=1/2$ exceptional eigenvalues due to the divergence in the $G$-functions.
For the present model parameters, no exceptional solution is missed in the
stable zeros of the $G$-function. In principle, the condition for the
occurrence of the exceptional eigenvalues is hardly satisfied for given
rational model parameters.

Note that for the $N=3$ Dicke model, the present $G$-function within ECS is only
a $2\times 2$ determinant, much simpler than $G$-functions with $6\times 6$
determinant derived by Braak using the Bargmann representation~\cite
{Braak2013}. The $G$-curves are different for the same model parameters, but
the stable zeros should be the same. What is more, we can study the large $N$
Dicke model straightforwardly. The only thing we need do is changing the
value of $N$ in these formulae. We like to stress here that increasing the
number of qubits $N$ would almost not bring much more additional effort to
search for the zeros of the present $G$-function numerically.

\begin{figure}[tbp]
\center
\includegraphics[width=12cm]{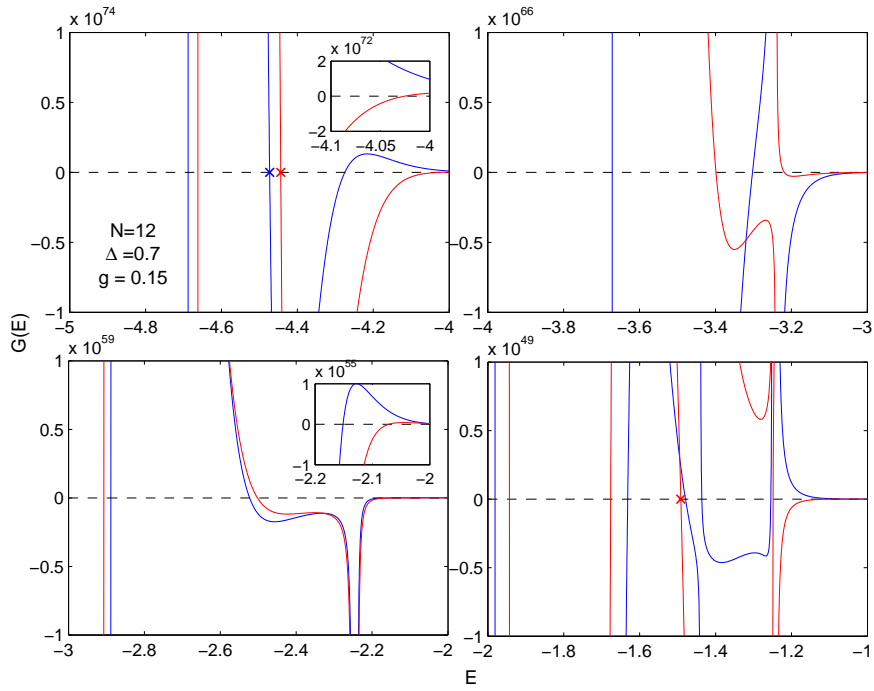}
\caption{ (Color online). $G$-functions for the $N=12$ Dicke model with
positive (blue) and negative (red) parity with different scales in four
energy intervals connected successively. Crosses denote the instable zeros. $%
\Delta =0.7$ and $g=0.15$. }
\label{G_function_N8}
\end{figure}

By $6\times 6$ determinant, we can study the $N=12$ Dicke model, which is
sufficient for the recent multi-qubits constructed on the superconducting
circuits for scalability. The corresponding $G$-function is exhibited in
Fig. \ref{G_function_N8} for $\Delta =0.7$ and $g=0.15$, equivalently $%
\lambda \approx 0.52$, slightly higher than the critical coupling. The
unstable zeros can be also easily excluded by the same procedure as stated
above. To reduce the magnitude of the value of $G$-function in the large $N$
Dicke model, in practice, one can re-scale the elements in Eq. (\ref
{G-function}) in each row by a same real number. The shape of the $G$
function will be modified, but the positions of the stable zeros remain
unchanged.

\begin{figure}[tbp]
\center
\includegraphics[width=10cm]{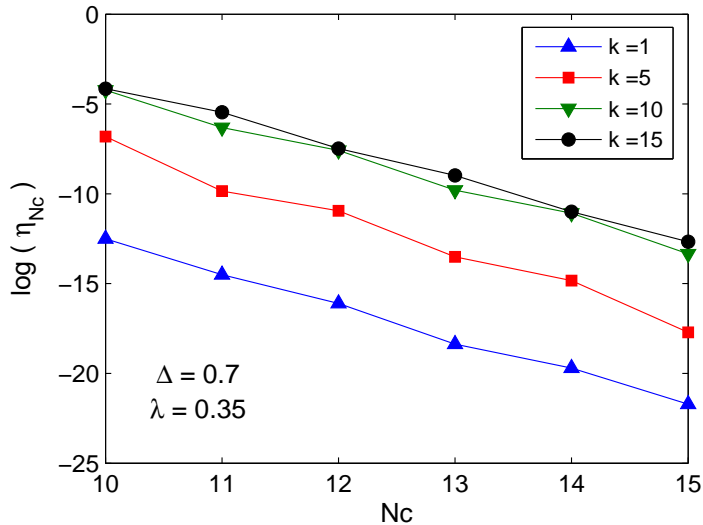}
\caption{ (Color online). The energy relative difference  as a function of the truncation number $N_c$ for  several eigenstates  in the $N=12$ Dicke model. $k$ numbers the eigenstates from the ground states $k=1$.
 $\Delta =0.7$ and $\lambda=0.35$. }
\label{compare_en}
\end{figure}

It is interesting to link coefficients in wavefunction (\ref{wavefunction})
in previous numerically exact techniques \cite{chen2008} and the present
wavefunctions (\ref{wave_A}), (\ref{wave_A-}), and (\ref{wave_d}) in the
same Dicke state $\left| \frac N2,m\right\rangle \;$
\begin{equation}
u_{m>0,n}\varpropto \sqrt{n!}c_{m>,n}^{(m)},\;\;u_{m<0,n}\varpropto \left(
-1\right) ^n\sqrt{n!}c_{-m,n}^{(m)},\;u_{m=0,n}\varpropto \sqrt{n!}a_{0,n}.
\end{equation}
The third one only exists for $N$ even. So the convergency  with the
truncated number $N_c$ of the ECS in both approaches should display
the similar behavior in the practical calculations, although the
previous coefficients are obtained from numerical diagonalization
and the present ones by the zeros of the $G$-functions. To show the
convergence, we present the relative difference of the energy
$\eta_{N_c} =\left| \left( E(N_c)-E(N_c-1)\right) /E(N_c)\right| $
as a function of $N_c$ in Fig. \ref{compare_en} for $N=12$ model. It
is observed  that the energy can converge rapidly with $N_c$ to any
desired accuracy. Interestingly, $\eta_{N_c}$ curves almost saturate
for the high excited states, demonstrating  the common advantage of
the ECS technique. Generally $N_c=20$ is sufficient to achieve the
eigenvalues with very high accuracy for a large region of the model
parameter space, similar to the numerical diagonalization in the ECS basis~%
\cite{chen2008,Bastarrachea,Hirsch1,Hirsch2}. The magnitude of the value of $%
G$-function will increase rapidly with $N$ in the low energy region, so re-scaling and/or high
precision computations are needed in some cases. Actually, we do not have to
plot the $G$-functions to obtain the exact solutions and only locate all
zeros numerically for both $N_c$ and $N_c+1\;$instead, followed by checking
the stability of all zeros.

\textsl{Non-integrability}: Both Dicke and Rabi models possess the $Z_2$ symmetry and have therefore
two conserved quantities, energy and parity. The dimension of the
finite-dimensional factor of the Hilbert space is $N+1$ in the Dicke case,
thus greater than two for $N>1$, while in the Rabi model it is two. The
parity symmetry provides two labels and this suffices to label each
state uniquely for $N=1$ but not for $N>1$. It follows that the Dicke model
is non-integrable \cite {Braak2013}. It is, however, analytically solvable, as shown
in the previous section. This demonstrates again (another case is the
driven Rabi model ~\cite{Braak}) that integrability and solvability are different
concepts, at least concerning systems with constituents in the deep
quantum limit.  Braak's criterion is consistent with both
the manifestation of chaos in the semiclassical counterpart of the
Dicke model~\cite{nonint} and the ''picket-fence'' character of
spectrum in the Rabi model~\cite{kus}.

By the structure of the derived $G$-functions, we will discuss the
non-integrability of the finite Dicke model in terms of the energy level
distribution in some detail below.

First, we review two limits of the Dicke model. In the limit of
$g=0$, the atoms are decoupled from the field, the eigenvalues are
\begin{equation}
E_{m^{\prime },n^{\prime }}=n^{\prime }+\Delta m^{\prime },  \label{g0}
\end{equation}
where $n^{\prime }\left( =0,1,2,...\infty \right) \ $ is the
photonic number in the original Fock space $d^{\dag }$, $m^{\prime
}\left( =-j,...,+j\right) $ is the eigenvalues of $J_x$. In this
case, the system is certainly integrable. The symmetry is not weaker
than that  in the rotating-wave approximation~\cite{RWA}.

In the strong coupling limit $g\rightarrow \infty $, the off-diagonal
elements in (\ref{Hamiltonian}) can be neglected, the system can be
described by the direct product of the Dicke state $\left| \frac
N2,m\right\rangle $ and number states in the displaced operators $A_m^{\dag
} $, the eigenvalues are then given by
\begin{equation}
E_{m,n}=n-4m^2g^2,  \label{g_infty}
\end{equation}
where $n\;\left( =0,1,2,...\infty \right) \ $ is the photonic number in $%
A_m^{\dag }$, $m\;\left( =-j,...,+j\right) $ is the eigenvalue of
$J_z$. They are  doubly degenerate for $m\neq 0$. The whole spectrum
must show a regular behavior. Interestingly they have the same
expressions as the poles of the $G$-functions at finite coupling.
Based on  the closed-form solution (\ref{g_infty}), the finite $N$
Dicke model in this limit is considered to be integrable by Emary
and Brandes~\cite{quntnoint,Emary}. The degenerate atomic limit
$\Delta =0$ is actually equivalent to the strong coupling limit.
Recently, Batchelor and Zhou argued that both Dicke and Rabi models
are integrable in this limit~\cite{Batchelor} in the sense of
Yang-Baxter concept~\cite{Yang-Baxt}.

\begin{figure}[tbp]
\center
\includegraphics[width=7cm]{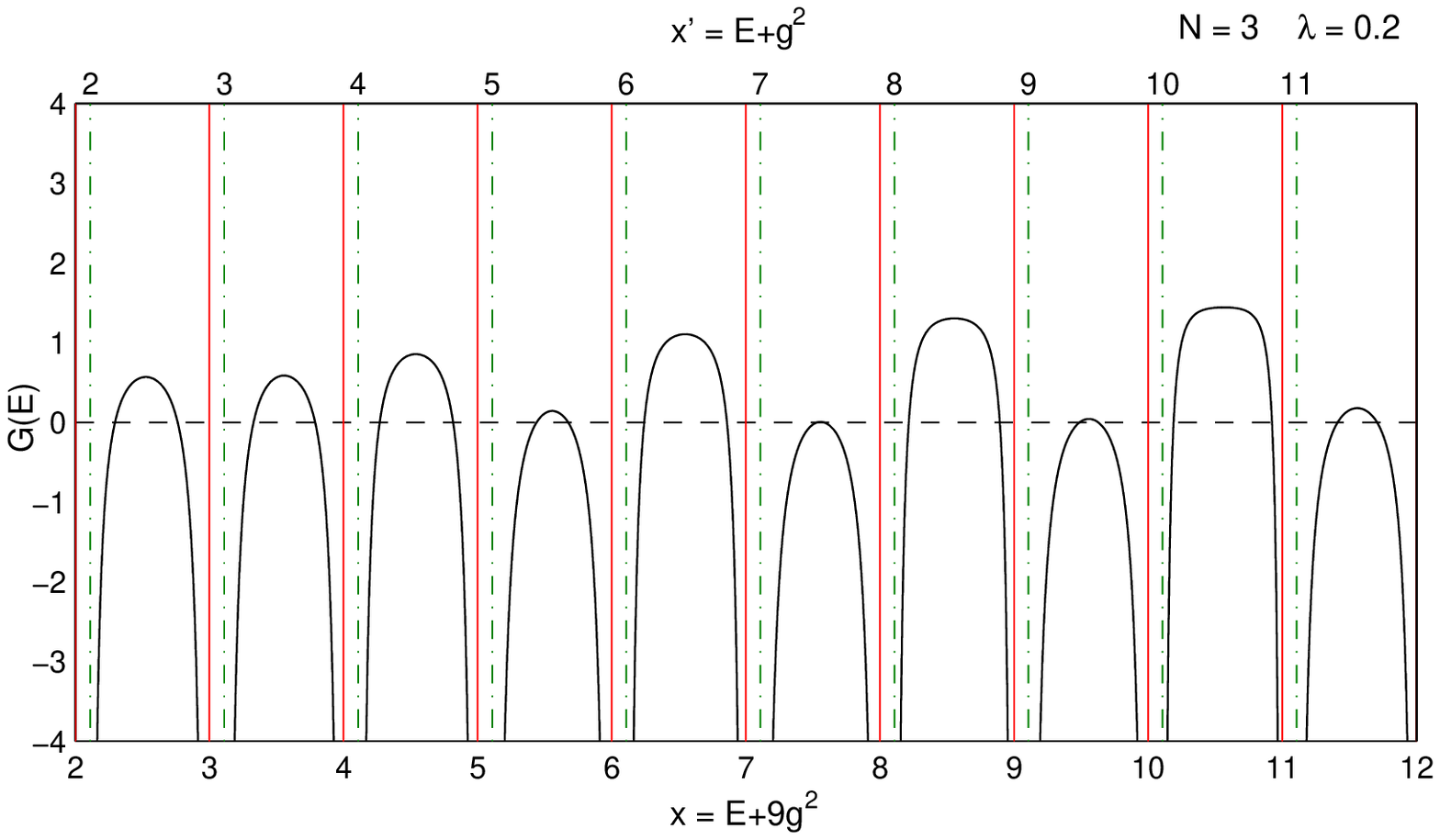} %
\includegraphics[width=7cm]{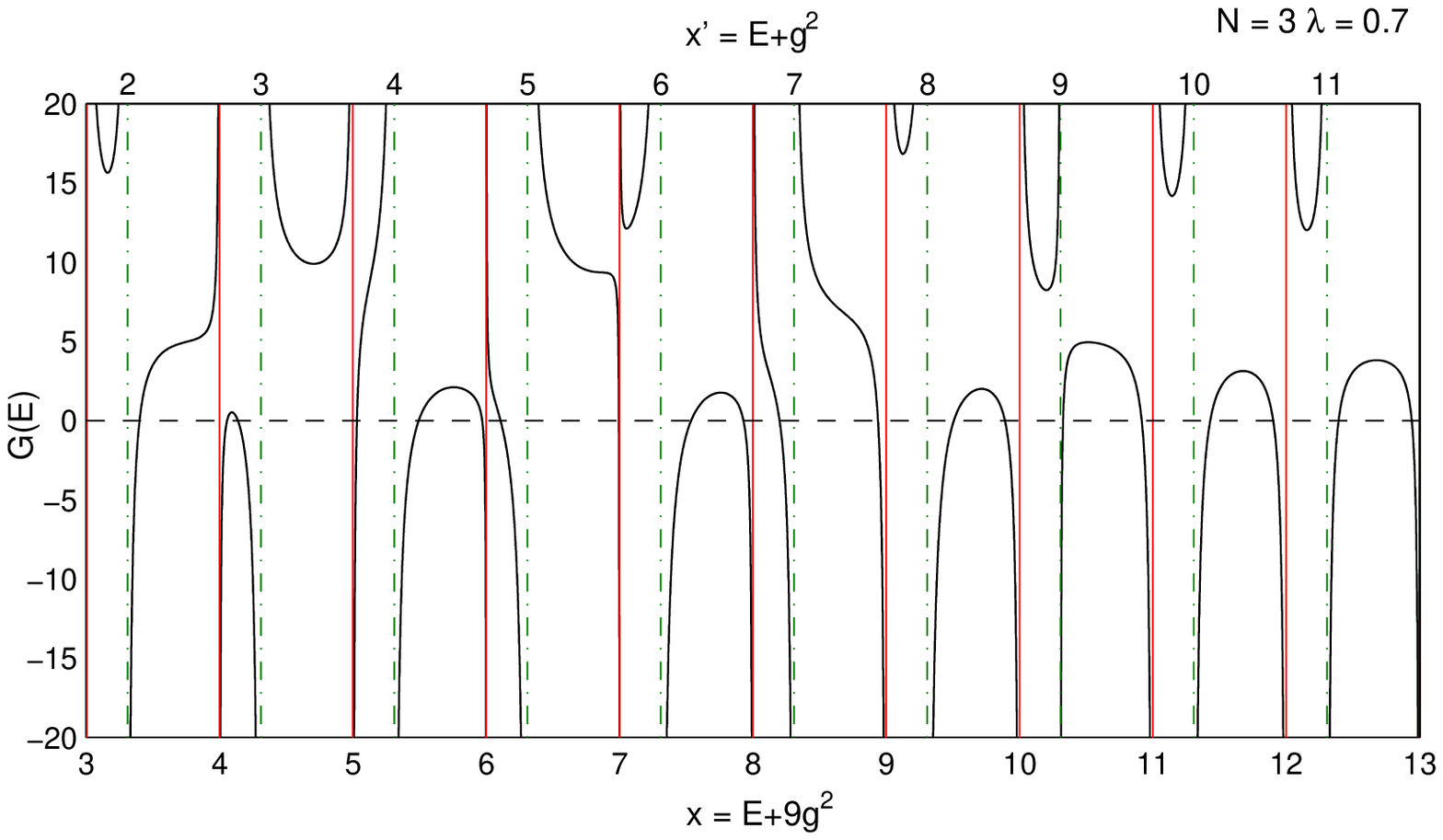} %
\includegraphics[width=7cm]{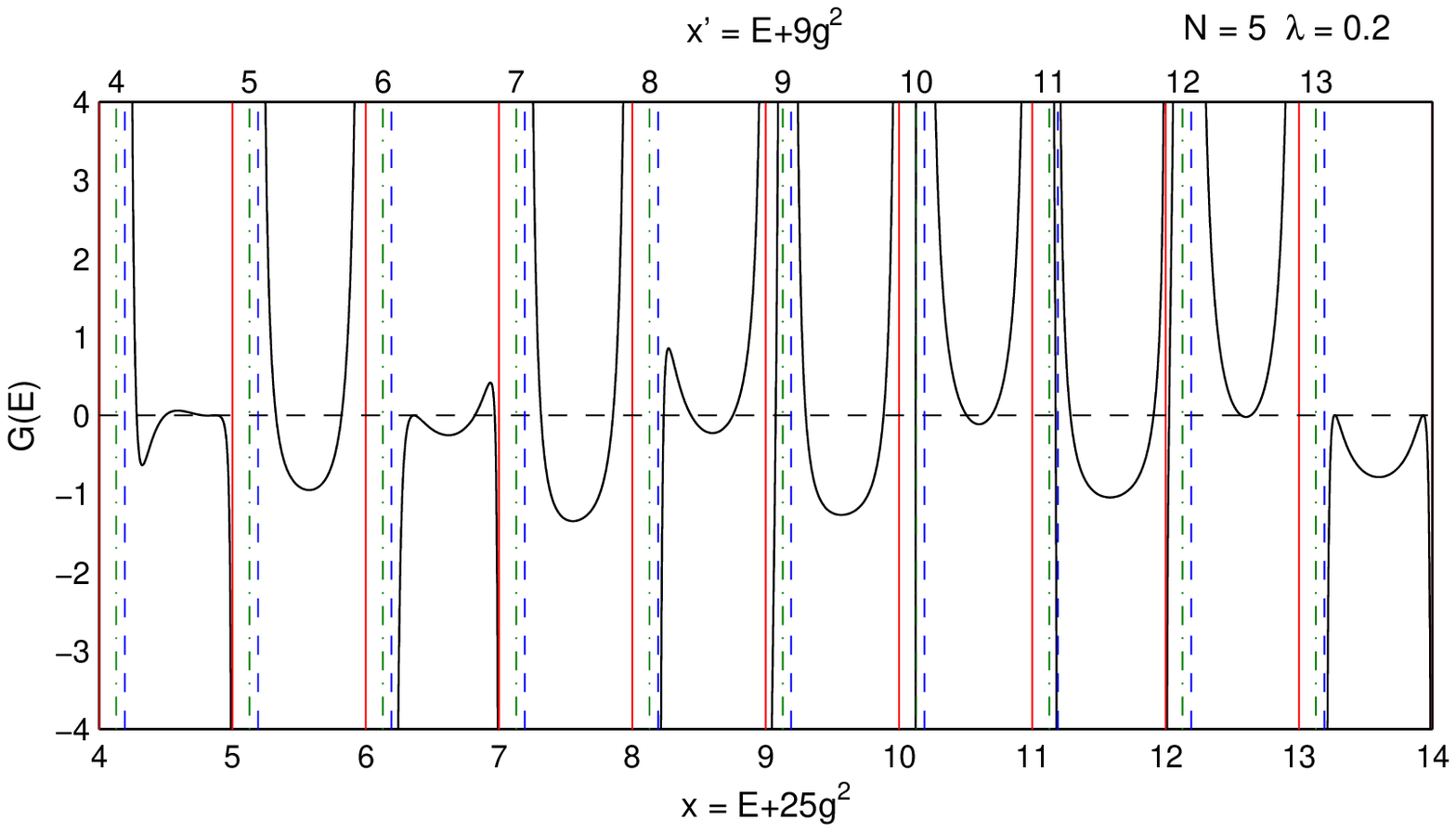} %
\includegraphics[width=7cm]{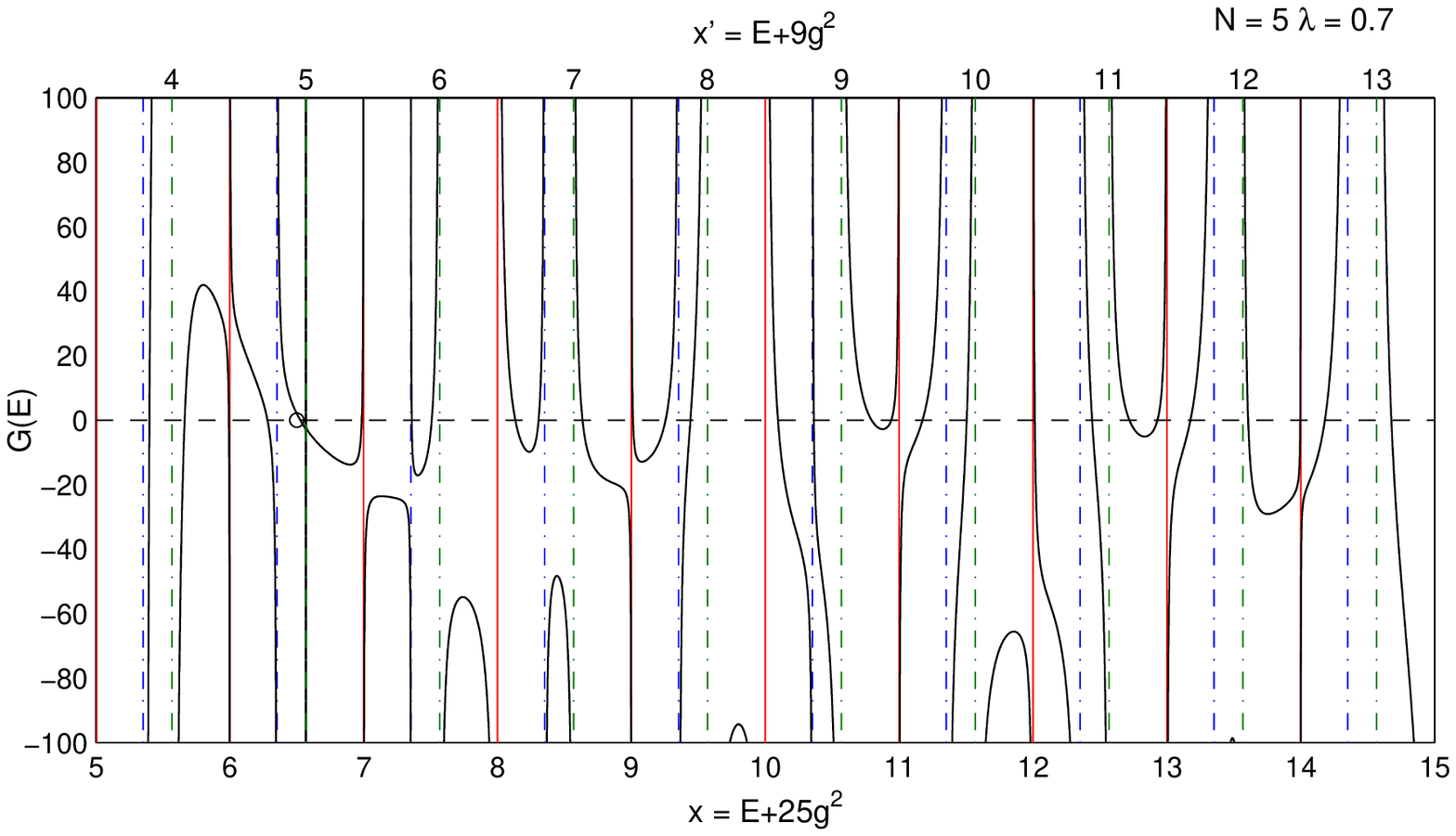}
\caption{ (Color online). $G$-functions with positive parity (solid black)  for the $N=3$ (upper) and $5$
(lower) Dicke models at  $\lambda=0.2$  (left) and $0.7$ (right). $x=n$ and $x^{\prime}=n$
baselines are denoted by red and green lines, and third baselines for $%
E+g^2=n$ of $N=5$ model by blue line. Open circles denote the unstable
zeros. $\Delta =1$. }
\label{G_function_N35}
\end{figure}

Generally, there are $N+1$ eigenvalues in the unit energy interval
in the weak coupling limit and $\frac{N+2}2\left( \frac{N+1}2\right)
$ eigenvalues for $N$ even (odd) in the strong coupling limit in the
$E>0$ region. The nontrivial eigenvalues at  finite coupling should
interpolate from two limits given by Eqs. (\ref{g0}) and
(\ref{g_infty}) which can be located by the zeros of $G$-function.
Without loss of generality and also for brevity, we demonstrate the
$G_{+}\left( x\right) $-functions, where $x=E+N^2g^2$, for $N=3\
$and $5$ in Fig. \ref{G_function_N35} for two typical coupling
strengths below and above $\lambda _c=1/2$ to explain the
non-trivial level
statistics. The baselines due to divergence are at the poles $%
x_n^{(m)}=n+\left( N^2-4m^2\right) g^2$. The $G $-function is not
analytic at the baselines, so the positions of the zeros are
dictated by the pole structure of $G$-function. It is clearly
exhibited in Fig. \ref{G_function_N35} that all segments of the
$G$-curves within the same parity start from one baseline and end at
the adjacent baseline, indicating that the zeros must be restricted
in the subintervals given by the nearest neighbor baselines. In this
way, the level distribution is dominantly controlled by the pole
structure of $G$-functions.

For small coupling, the baselines $x_n^{(m)}\;$within the same $n$
are very close, so zeros in each parity subspace in the interval
$\left[ n,n+1\right] $ prefer to stay in the wider subintervals on
the right side rather than the narrow subintervals on the left side.
As $N$ increases, one can conjecture that the zeros are squeezed to
few wider subintervals, leading to relatively more very small
avoided crossings exhibited in the spectral graph. Interestingly,
more small avoiding crossings for $N=5$ than $N=3$ are obviously
seen in Fig. \ref {G_function_N35} at $\lambda =0.2$. The general
trend can be confirmed numerically.  Note that  the present
$G$-function is single valued for the energy variable, so only the
level repulsion is allowed and the level crossing can be excluded in
each parity subspace. Although we could not rule out the rare events
that the $G$ curve is exactly tangent to $G=0$ line if the extremal
condition is also met, leading to the level "collisions" in the
spectral graph, they are however only related to fragile
degeneracies and by no means the true level crossing. There is no
visible difference between a true crossing and a small avoided
crossing in the level statistics, so the Poissonian statistics
observed at the weak coupling~\cite{Emary} is a consequences of
rather many very small avoided crossings than many true level
crossings. The level crossing is forbidden by non-integrability. The
Berry-Tabor criterion may only be suited to the quantum models
having the classical limits, which comprise many important systems
including the infinite $N$ Dicke model~\cite{Emary}, but not the
finite Dicke model.

As the coupling increases, the largest  subinterval between $x_n^{(N/2)}$and
$x_n^{(N/2-1)}$ can exceed the main interval $\left[ n,n+1\right] $. In this
way, the crossover coupling constant can be estimated as
\[
\lambda _c^{(N)}=\frac 12\sqrt{\frac N{N-1}}.
\]
Very interestingly, for large $N$, $\;\lambda _c^{(N)}$ is close to$\ 1/2$,
the critical point of the quantum phase transitions. For finite but large $N$%
, many wide subintervals emerge above $\lambda _c^{(N)}$, and then provide
the substantial ways to assign the zeros, allowing an irregular distribution of
the energy level. This trend can be also seen in Fig. \ref{G_function_N35}
at $\lambda =0.8$. Alleviating the jam of the zeros in few subintervals
would reduce the avoiding crossing and then lead to the Wigner-like
statistics at strong coupling, consistent with the observation in the level
statistics~\cite{Emary}. Specially, for the Rabi model, there is no
subinterval and the only interval $\left[ n,n+1\right] $ remains unchanged
for arbitrary coupling, so the energy distribution is more regular than that
in the Dicke model~\cite{kus, Braak}.

To this end, we believe that the Dicke model is exactly solvable but not
integrable for any finite $\lambda $ and $N>1$. Besides these fundamental
issues, we may also find many practical applications of the present
analytical exact technique in the physical problems. In the next section, we
will apply it to the exact dynamics of GME of this model, which has not been
explored before.

\section{Dynamics of genuine multipartite entanglement}

First, we briefly review the scheme to the calculation of the GME. A state $%
\rho $ for multipartite systems is called biseparable if it can be written
as a mixture of states separable with respect to different bipartitions. It
is well-known that separable states are always positive partial transpose
(PPT)~\cite{Peres}, indicating that a set of separable states with respect
to some bipartition is contained in a larger set of states with PPT for the
same partition. Since any biseparable state is a PPT mixture, a state which
is not PPT mixture implies genuine multipartite entanglement.

Recently, a new approach to characterize genuine multipartite entanglement
has been proposed~\cite{Jungnitsch} by considering the PPT mixture. PPT
mixtures can be characterized by the method of semidefinite programming~\cite
{Vandenberghe}. Given a state $\rho $ for multipatite systems, one can
search the minimization $\ $ of the trace of matrix $W\rho ,$ where $W$ is a
decomposable entanglement witness for any bipartition. It has been proved
that a state is PPT mixture if minimum of $Tr(W\rho )$ is positive. If the
minimum is negative, then $\rho $ is not a PPT mixture hence is genuinely
multipartite entangled. The optimization problem can be solved by using
YALMIP~\cite{YALMIP}, SEDUMI~\cite{SEDUMI} or SDPT3~\cite{SDPT3}, a
ready-to-use freely available implementation. Moreover, the absolute value
of such minimum (denoted by$E(\rho )$) was proved to be an entanglement
monotone for genuine multipartite entanglement~\cite{Jungnitsch}.

In the finite $N$ Dicke model, we start from the maximum entangled
states in the $N$-qubit basis
\[
\left| N\right\rangle _{e_{\max }}=\left( \left| \uparrow \uparrow
,...,\uparrow \right\rangle _N+\left| \downarrow \downarrow ,...\downarrow
\right\rangle _N\right) /\sqrt{2},
\]
which are just the Bell states for $N=2$ and GHZ states for $N>2$. In the
basis of Dicke states, it can be rewritten as
\begin{equation}
\left| N\right\rangle _{e_{\max }}=\left( \left| \frac N2,\frac
N2\right\rangle +\left| \frac N2,-\frac N2\right\rangle \right) /\sqrt{2}.
\label{initialstate}
\end{equation}
The initial field state is the vacuum state $\left| 0\right\rangle _d$.
Since $[H,J^2]=0$, the state will only evolve in the subspace spanned by the
basis $\left| \frac N2,m\right\rangle \bigotimes \left| l\right\rangle _d$.
Using the $n$-th eigenvector of the Dicke model given in Sec. II
\begin{equation}
\left| n\right\rangle _{}=\sum_{m=-N/2}^{N/2}|\frac N2,m\rangle |\phi
_m\rangle _d,
\end{equation}
we can expand initial state (\ref{initialstate}) as
\begin{equation}
\left| N\right\rangle _{e_{\max }}=\sum_nf_n\left| n\right\rangle .
\end{equation}
The evolution of wave function thus is given by:
\begin{equation}
|\psi (t)\rangle =e^{-iHt}\left| N\right\rangle _{e_{\max
}}=\sum_ne^{-iE_nt}f_n\left| n\right\rangle .
\end{equation}

\begin{figure}[tbp]
\center
\includegraphics[width=12cm]{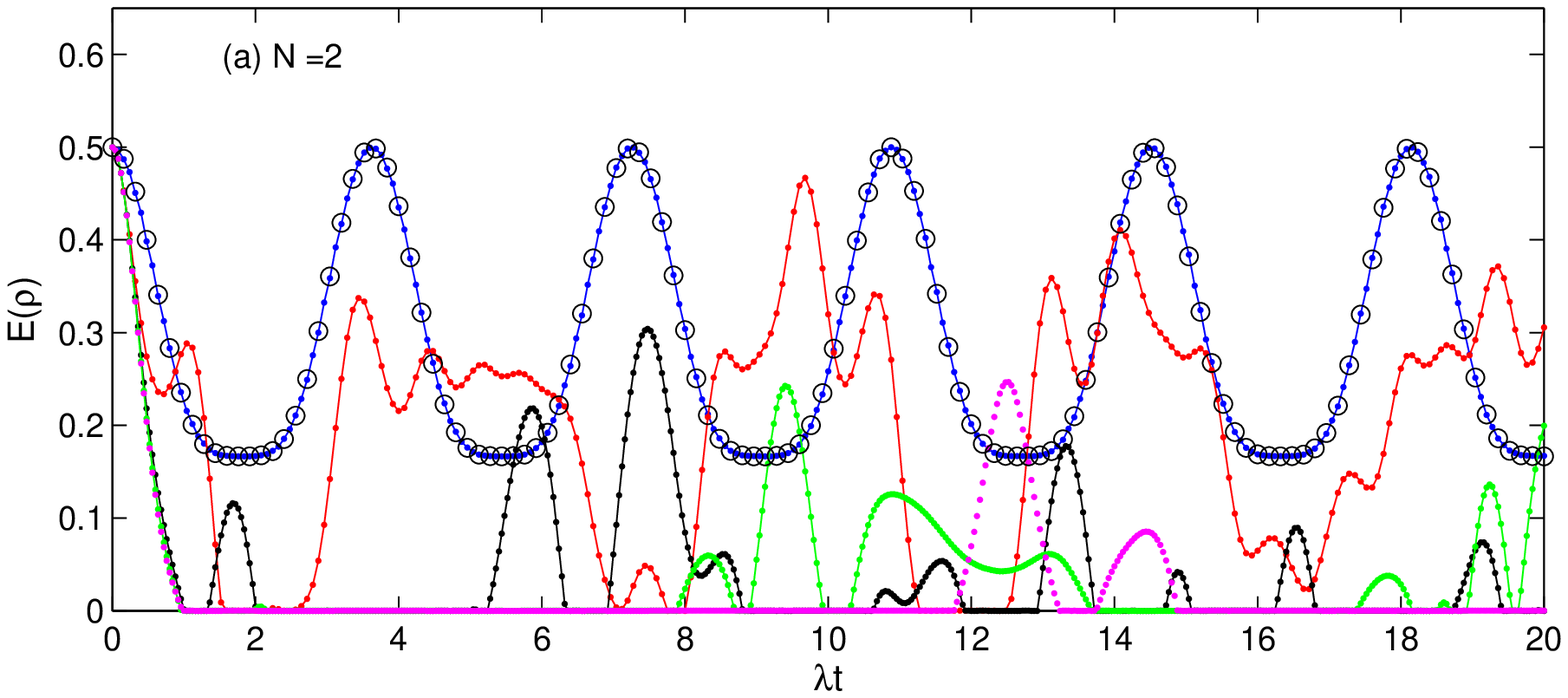} %
\includegraphics[width=12cm]{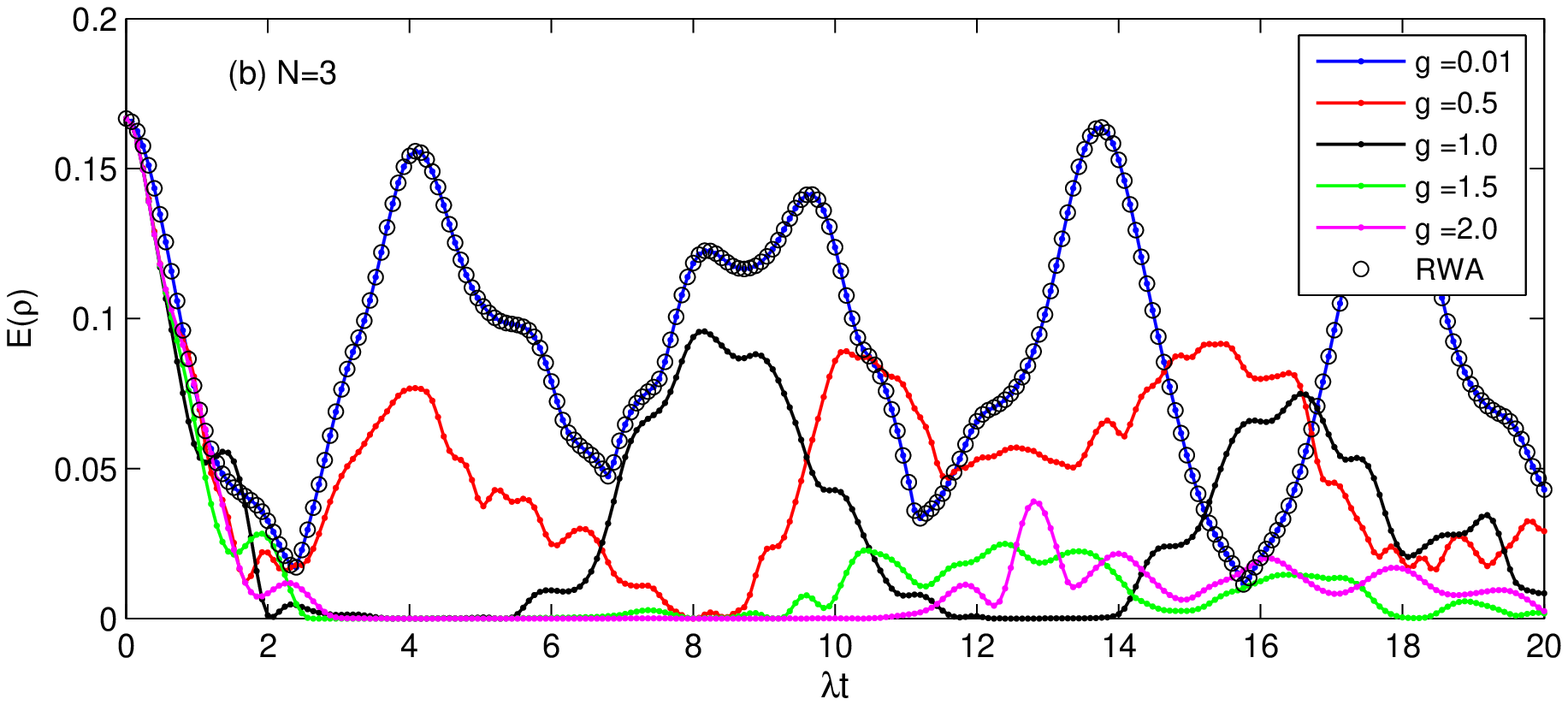} %
\includegraphics[width=12cm]{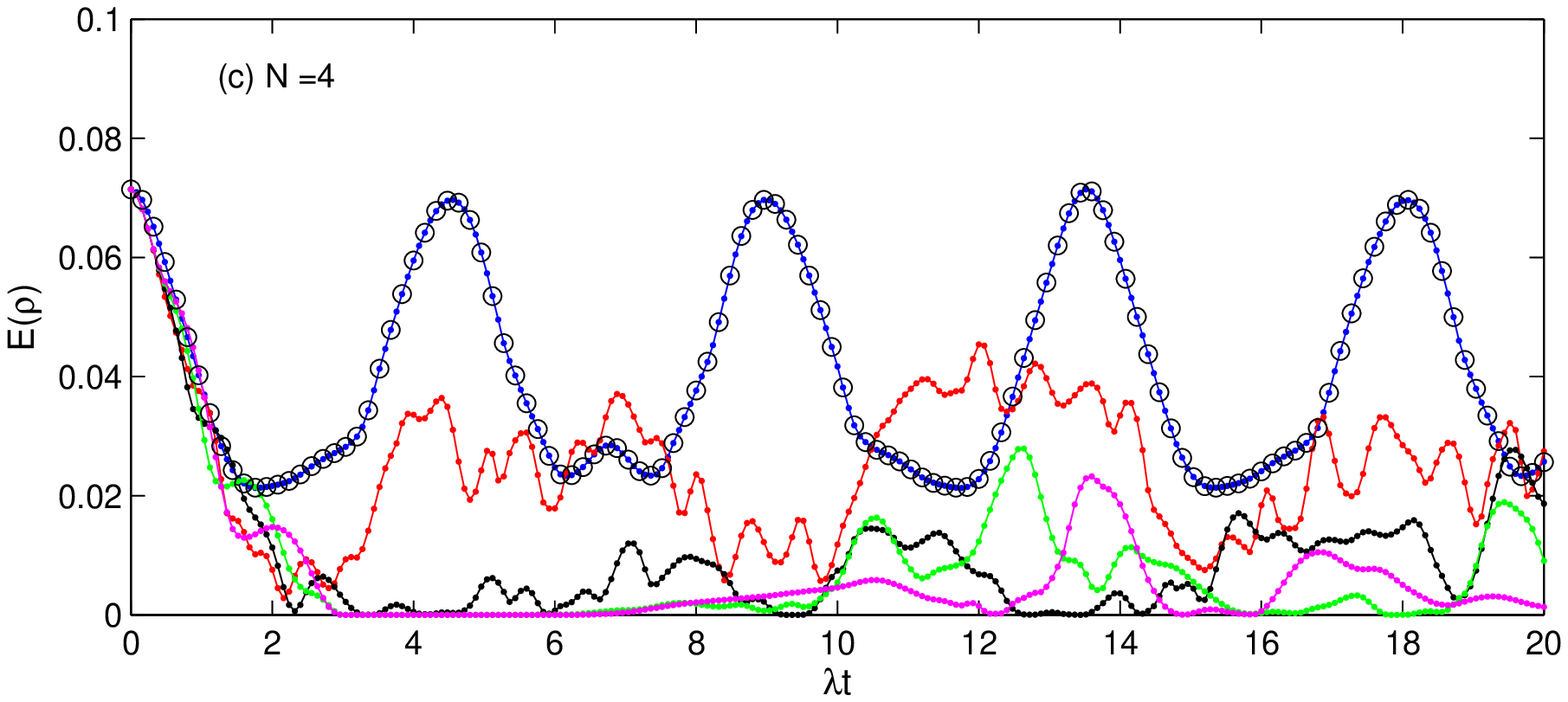}
\caption{ (Color online) The evolution of the GME from the maximum entangled
state for different coupling strengths denoted with different colors in $N=2$
(a), $N=3$ (b), and$N=4$ (c) Dicke models. $\Delta =1$ }
\label{dynamics_N234}
\end{figure}

To calculate the dynamics of the GME, each Dicke state $|\frac N2,m\rangle $
should be expressed in $N$-qubit basis$\;|j\rangle _q\;$($j=1,...,2^N$). The
atomic reduced density matrix $\rho $ can be calculated by tracing out the
photonic degrees of freedom $\rho (t)=$Tr$_{photon}|\psi (t)\rangle \langle
\psi (t)|$ with dimension $2^N$.

The evolution of the GME initiated from the qubit maximum entangled state
and field vacuum state can be studied within the above definition and the
exact eigensolutions. Fig. \ref{dynamics_N234} presents the GME dynamics for
different coupling constant $g$ at resonance $\Delta =1$ for $N=2,3$, and $4$
respectively. Periodic or quasi periodic behavior for the GME dynamics is
observed at the very weak coupling, which is basically the same as the
results in the RWA where the photonic number is conserved. With the
increasing coupling, the regular behavior is gradually destroyed, mostly due
to the activation of more photons. It is worth further study whether the
''chaotic'' behavior at the strong coupling is relevant with the quantum
chaos in this model~\cite{Emary}. Very interestingly, we find that at the
same coupling strength ($g$) between the single-mode cavity and the
individual qubit, the GME for more qubits are more stronger, which will be
an advantage if GME for more qubits act as the resource in the quantum
information processing.

\section{Summary}

In this work, we have derived a concise $G$-function by a $k\times k$
determinant for the arbitrary $N=2k-1$ and $2k$ Dicke model.
This $G$-function is a well defined transcendental function, so formally
analytical exact solutions to the Dicke model is found in a strict
mathematical sense. Without built-in truncations, the present method is
essentially different from the previous numerical exact ones, therefore of
more academic values. The mathematics behind the $G$-function is very
interesting and may be worth further exploration in the future. This work is
to extend the methodology of $G$-function in the Rabi model to the identical
multi-qubit cases, thereby allowing not only in-depth studies in some
fundamental issues but also practically feasible treatment to energy spectra
and eigenstates.

The stable zeros of this $G$-function will lead to exact eigensolutions of
the model, which are useful for many applications in the Dicke model. The
GME dynamics has been calculated for Dicke model with a few number of qubits  as a
example in this paper. It is shown that GME dynamical behavior is strongly $%
N $ dependent. The GME becomes stronger with the increasing qubit number for
the same coupling.

In the Dicke model, we find that the exact regular spectrum can be described
in terms of infinite polynomials, and can be simply located numerically. In this sense,
one may argue that these exact solutions are not fully analytic.
While the isolated exact exceptional ones
can be expressed algebraically, actually a parabolic function of $g$. In any case, the
exact solvability has been  demonstrated without doubt. The eigenstates
cannot be uniquely specified by the quantum numbers representing the
continuous and discrete d.o.f, suggesting non-integrability of the $N>1$
Dicke model at any finite $\lambda $ in terms of Braak's criterion. The
level distribution is mainly controlled by the pole structure of the $G$%
-functions, which is determined by the exceptional spectrum. We
argue that the Poissonian statistics at the weak coupling observed
in literature is the consequence of rather many very small avoided
crossings than many true level crossings. The absence of the more
level crossings than allowed by parity symmetry is a strong
indication of non-integrability. The present work may add a new
example to the non-integrable but exactly solvable models.

\textbf{Acknowledgments}

QHC thanks Daniel Braak for stimulating and illuminating discussions and the detailed
explanation of his criterion for integrability. This work is supported by
National Basic Research Program of China, 2011CBA00103, National Natural
Science Foundation of China, 11174254, 11474256.


\end{document}